\documentclass[11pt,a4paper]{article}
\pdfoutput=1
\usepackage{
	jheppub,
	amsthm,
	amscd,
	amsfonts,
	bbold,
	MnSymbol
}
\usepackage[normalem]{ulem}
\usepackage{tocloft}

\def\a{\alpha} \def\b{\beta} \def\g{\gamma} \def\d{\delta} \def\e{\epsilon} \def\f{\phi}  \def\ve{\varepsilon} \def\z{\zeta} \def\h{\eta} \def\k{\kappa} \def\l{\lambda} \def\m{\mu} \def\n{\nu}  \def\r{\rho}  \def\s{\sigma} \def\t{\tau}  

       \def\Th{\Theta}    \def\L{\Lambda}         \def\W{\Omega}

\def\fr{\frac}  \def\dt{\partial}

\def\mc{\mathcal}  \def\mE{\mathcal{E}} \def\mF{\mathcal{F}} \def\mL{\mathcal{L}} \def\mR{\mathcal{R}}      \def\DD{{\mathcal{D}}} 
\def\XX{\mathbb{X}} \def\RR{\mathbb{R}} \def\SS{\mathbb{S}} \def\bg{\bar{g}}

\title{Non-abelian tri-vector deformations\\ in $d = 11$ supergravity}

\author[a,b]{Ilya Bakhmatov,}
\author[c]{Kirill Gubarev,}
\author[c,d]{Edvard T. Musaev,}

\affiliation[a]{Institute for Theoretical and Mathematical Physics, Lomonosov Moscow State University, Lomonosovsky avenue, Moscow, 119991, Russia}
\affiliation[b]{Yerevan Physics Institute, Alikhanian Brothers 2, Yerevan, 0036, Armenia}
\affiliation[c]{Moscow Institute of Physics and Technology, Institutskii per. 9, Dolgoprudny, 141700, Russia}
\affiliation[d]{Kazan Federal University, Institute of Physics, Kremlevskaya 16a, Kazan, 420111, Russia}

\emailAdd{ibakhmatov@itmp.msu.ru}
\emailAdd{kirill.gubarev@phystech.edu}
\emailAdd{musaev.et@phystech.edu}

\abstract{A truncation of the SL(5) Exceptional Field Theory that allows to describe spacetimes of the form $M_4 \times M_7$ with the 4-form flux on $M_4$ is constructed. The resulting theory is used to test the recently proposed tri-vector generalisation of Yang-Baxter deformations applied to the AdS${}_4 \times \SS^7$ solution of $d=11$ supergravity. We present two new supergravity solutions corresponding to non-abelian non-unimodular tri-vector deformations of AdS${}_4 \times \SS^7$.}

\keywords{Supergravity Models, Space-Time Symmetries, String Duality}

\arxivnumber{2020.01915}

\begin{document}

\maketitle

\section{Introduction}

Supergravity interpretation of integrable deformations of string theory $\sigma$-models has seen rapid progress in the recent years. Yang-Baxter deformations~\cite{Delduc:2013fga,Delduc:2013qra,Kawaguchi:2014qwa}, $\h$-deformed~\cite{Klimcik:2002zj,Klimcik:2008eq} and $\l$-deformed~\cite{Sfetsos:2013wia,Hollowood:2014qma} $\s$-models may all be represented by combinations of T-dualities~\cite{vanTongeren:2015soa,Osten:2016dvf}, as well as their non-abelian~\cite{Hoare:2016wsk,Borsato:2016pas,Borsato:2017qsx} and Poisson-Lie~\cite{Klimcik:2015gba} extensions. An element of the T-duality group $O(d,d)$, acting on a supergravity background, can be conveniently represented by the so-called $\b$-shift, parametrised by a bi-vector $\b$ \cite{Lust:2018jsx, Sakamoto:2018krs}. Basic building blocks of integrable deformations in the supergravity language, the Lunin-Maldacena (TsT)~\cite{Lunin:2005jy,Frolov:2005dj} transformations, correspond to constant $\b$~\cite{CatalOzer:2005mr}. General Yang-Baxter deformations result from using an $r$-matrix solution to the classical Yang-Baxter equation as a deformation bi-vector. The transformation of the NSNS supergravity background fields $(g,b) \to (G,B)$ is given by the Seiberg-Witten open/closed string map~\cite{Seiberg:1999vs}, with $\b$ playing the role of an anticommutativity parameter~\cite{Araujo:2017jkb,Araujo:2017jap}. Extended to the case of backgrounds with the $b$-field, this map takes the form:
\begin{equation}
\label{beta_map}
      G + B = (g + b) \big(1 + \b (g + b)\big)^{-1}.
\end{equation}
Here $\b = \fr12 r^{\a\b}k_\a \wedge k_\b$ is the deformation bi-vector written in terms of a constant antisymmetric $r$-matrix and the Killing vectors of the initial background, which obey the isometry algebra $[k_\a, k_\b] = f_{\a\b}{}^\g k_\g$. Assuming that $r^{\a\b}$ satisfies the classical Yang-Baxter equation,
\begin{equation}
\label{cybe}
    r^{\a[\g} r^{|\b|\d} f_{\a\b}{}^{\e]} =0,
\end{equation}
is sufficient for the deformed fields $G,B$ to be a supergravity solution~\cite{Borsato:2018idb,Bakhmatov:2018bvp}. This allows to view the map~\eqref{beta_map} as a supergravity solution generating method, valid for generic spacetimes with isometries~\cite{Bakhmatov:2017joy,Bakhmatov:2018apn}.

The reason that the classical Yang-Baxter equation~\eqref{cybe} is instrumental in the $d=10$ deformation prescription is ultimately that the two-dimensional string worldsheet theory exists behind the scenes of the supergravity approximation. Similarly, it is natural to expect that some fundamental properties of M-theory could be manifested by finding a consistent extension of the Yang-Baxter deformations to the $d=11$ supergravity. In the absence of an M-theory version of the $\s$-model deformation narrative, we propose that supergravity symmetries can be employed to build such a generalisation. 

Supergravity formulations that are natural to look at in this context are the Double~\cite{Hull:2009mi} and Exceptional~\cite{Hohm:2013pua} Field Theories (DFT and ExFT, respectively). Specifically designed to render supergravities in various dimensions covariant under T- and U-duality groups at the expense of extending the spacetime dimension, they are useful in describing Yang-Baxter deformations~\cite{Sakatani:2016fvh,Baguet:2016prz,Sakamoto:2017cpu,Fernandez-Melgarejo:2017oyu,Sakamoto:2018krs} and Poisson-Lie T-duality~\cite{Hassler:2017yza,Lust:2018jsx,Demulder:2018lmj,Sakatani:2019jgu,Demulder:2019bha,Sakatani:2019zrs,Malek:2019xrf}. The proof of~\cite{Bakhmatov:2018bvp} that~\eqref{beta_map},~\eqref{cybe} is a supergravity symmetry relied upon the DFT techniques, in particular the $\b$-supergravity formalism~\cite{Andriot:2011uh, Andriot:2013xca, Andriot:2014uda}. In this approach the map~\eqref{beta_map} is viewed as an expression of the intrinsic freedom of frame choice in DFT, which admits a straightforward extension into the ExFT realm, and hence to $d=11$ supergravity. The deformation bi-vector $\b$ becomes a dynamical field, and it can be shown that the CYBE~\eqref{cybe} is sufficient to put $\b$ on-shell.

In~\cite{Bakhmatov:2019dow} a tri-vector deformation prescription for $d=11$ supergravity was proposed, based on the freedom of frame choice in the SL(5) ExFT~\cite{Blair:2014zba}. The NSNS 2-form $b$ and the deformation bi-vector $\b$ are replaced by rank 3 tensors $C$ and $\W$, with a Killing tri-vector ansatz for the latter,
\begin{equation}\label{3k}
    \W = \frac{1}{3!} \r^{\a\b\g} k_\a \wedge k_\b \wedge k_\g,
\end{equation}
and a slightly more involved deformation prescription~\eqref{Cdef} instead of the open/closed map. When the Killing vectors form a U(1)$^3$ subgroup, this prescription reproduces an uplift of TsT to $d=11$~\cite{CatalOzer:2009xd,Deger:2011nb}. For a single commuting $\dt_*$ the tri-vector can be written as $\W = \dt_* \wedge \b$, and non-abelian YB deformations with respect to $\b$ can be recovered after the dimensional reduction. Whether any intrinsically 11-dimensional deformations exist has been left an open question because of a technical restriction imposed by the formalism. Namely, the simplified SL(5) ExFT setup of~\cite{Bakhmatov:2019dow} required that there be no flux of the 3-form $C$ inside the 4-dimensional submanifold, where the deformation acts. Thus, the consideration was essentially restricted to the flat space or a sphere and it was hard to come up with an isometry algebra nontrivial enough to provide a completely non-abelian~$\W$.

It is one of the aims of the present paper to overcome these restrictions. We adopt the approach similar to that of~\cite{Bakhmatov:2018bvp}, only now instead of the $\b$-supergravity field equations one has to deal with the dynamical equations of a certain truncation of the SL(5) ExFT. This theory is designed to describe the mechanics of U-duality within the 4-dimensional submanifold in a $4+7$ split. Thus, it can become a natural tool in studying tri-vector deformations of AdS${}_4$ within the Freund-Rubin solution. Conformal algebra of AdS${}_4$ is nontrivial enough to harbour non-abelian tri-vectors, so that the resulting deformations of AdS${}_4 \times \SS^7$ cannot be interpreted as mere uplifts of $d=10$ Yang-Baxter deformations in any obvious manner. 

Using the generators of momentum $P_a$, angular momentum $M_{ab}$, and dilatation $D$, we study the  deformations corresponding to $\W \sim P\wedge P\wedge M$ and $D\wedge P\wedge P$. Such $\W$ cannot be represented in the form $\W=\dt_{*} \wedge \b$  such that $\dt_*$ commutes with the generators of $\b$. More importantly, one shows that the 11-dimensional analogue of the $I$ vector of generalised supergravity is non-zero for these backgrounds. Hence, although these two deformed backgrounds are solutions of the conventional $d=11$ supergravity, one might expect them to connect to solutions of $d=10$ generalised supergravity upon reduction. This might be a hint of non-existence of an analogue to generalised supergravity in $d=11$.

The paper is structured as follows. After briefly introducing the SL(5) exceptional field theory in the Section~\ref{section:SL5}, we derive explicit relations between the fields of $d=11$ supergravity in the $4+7$ split and the ExFT. In the Section~\ref{section:eoms} we truncate the theory to backgrounds of the form $M_{11}=M_4\times M_7$ with the metric on $M_7$ that does not depend on coordinates of $M_4$. We define the deformation map for background with the 3-form flux on $M_4$ and provide equations of motion that the deformed background must satisfy. In the Section \ref{section:ads} we apply this formalism to the AdS${}_4\times \SS^7$ background and present the deformed solutions. We discuss the results in the Section \ref{section:conclusions}, and comment on the tentative $d=11$ generalisation of the CYBE that has yet to be determined.

\section{Exceptional field theory: SL(5) group}
\label{section:SL5}

The SL(5) exceptional field theory describes supergravity  dynamics, while being explicitly covariant under the transformations of the SL(5) U-duality group. The theory is formulated in terms of the fields
\begin{equation}
    \begin{aligned}
     & h_{\m\n}, && A_\m{}^{MN}, && B_{\m\n M}, && m_{MN},
    \end{aligned}
\end{equation}
that depend on 7 coordinates $y^\m$ parametrising the so-called `external' space and 10 coordinates $\XX^{MN}$ paramterising the so-called `internal' space ($M,N = 1,\ldots,5$ are fundamental SL(5) indices and $\XX^{MN}$ is antisymmetric. We refer to the appendix~\ref{section:app} for our index conventions). $h_{\m\n}$ is the external space metric, while $m_{MN}$, collection of scalars from the $d=7$ theory viewpoint, forms what is called the ExFT generalised metric. The internal space is endowed with the structure of extended geometry \cite{Berman:2010is,Berman:2011cg} with SL(5) as the group of local coordinate transformations. Under these a generalised vector $V^M$ of weight $\l$ transforms as defined by the following generalised Lie derivative
\begin{equation}
    \d_\L V^{M} = \fr12 \L^{PQ}\dt_{PQ}V^M -\fr14 (T^M{}_N)^{KL}{}_{PQ}\dt_{KL}\L^{PQ}V^N+\l \dt_{PQ}\L^{PQ}V^M,
\end{equation}
where $T^M{}_N$ represent the generators of SL(5). Algebra of such local generalised diffeomorphisms closes upon imposing the section constraint
\begin{equation}
    \e^{MNKLP}\dt_{MN} \bullet \dt_{KL} \bullet=0,
\end{equation}
where bullets represent any combinations of any fields. In what follows we will always assume the solution of the section constraint $\dt_{mn}=0$ that corresponds to $d=11$ supergravity~\cite{Musaev:2015ces} by removing the dependence on six out of ten extended coordinates ($m,n = 1,\ldots,4$).

Lagrangian of the SL(5) exceptional field theory reads \cite{Musaev:2015ces}
\begin{equation}
    \begin{aligned}
     e^{-1}\mL =& \ \hat{\mc{R}}[h_{(7)}] \mp \fr18 m_{MN}m_{KL}\mF_{\m\n}{}^{MK}\mF^{\m\n N L}+\frac{1}{48}h^{\m\n}\DD_{\m}m_{M N} \DD_\n m^{M N} + e^{-1}\mL_{sc}\\
     &+{\fr{1}{3\cdot (16)^2}}m^{MN}\mF_{\m\n\r M}\mF^{\m\n\r}{}_N+e^{-1}\,\mL_{top},
    \end{aligned}
\end{equation}
where $e=(\det h_{\m\n})^{\fr12}$ and the scalar part  part is given by
\begin{equation}
\label{Lsc}
    \begin{aligned}
      e^{-1} \mL_{sc}=& \pm \left(\frac{1}{8}\,  {\partial}_{MN}{{m}_{P Q}}\,  {\partial}_{KL}{{m}^{P Q}}\,  {m}^{M K} {m}^{N L} +\frac{1}{2}\, {\partial}_{MN}{{m}_{P Q}}\,  {\partial}_{KL}{{m}^{M P}}\,  {m}^{N K} {m}^{L Q}\right. \\
      &+ \frac{1}{2}\, {\partial}_{MN}{{m}^{L N}}\,  {\partial}_{KL}{{m}^{M K}} +\fr12 m^{MK}\dt_{MN}m^{NL}(h^{\m\n}\dt_{KL}h_{\m\n})\\
      &+ \left. \fr18 m^{MK}m^{NL}(h^{\m\n}\dt_{MN}h_{\m\n})(h^{\r\s}\dt_{KL}h_{\r\s}) +\fr18 m^{MK}m^{NL}\dt_{MN}h^{\m\n}\dt_{KL}h_{\m\n} \right).
    \end{aligned}
\end{equation}
Here and further in the text the upper sign corresponds to the case where the external $d=7$ space has Lorentzian signature, while the lower corresponds to the Euclidean signature. When the time direction falls into the internal space of ExFT (lower sign), we are dealing with timelike U-dualities~\cite{Hull:1998vg,Hull:1998ym}. The ExFT formalism relevant for this case has been developed in~\cite{Malek:2013sp}.
Irrespective of the signature choice, the U-duality group is SL(5); it is the local duality group that accommodates for the difference.

Splitting the fundamental SL(5) index as $M=1,\ldots,4,5=(m,5)$, components of the generalised metric can be parametrised as
\begin{equation}\label{mmetric1}
\begin{aligned}
			m_{MN}=h^{\fr{1}{10}}
				\begin{bmatrix}
				h^{-\fr12}h_{mn} && -V_{m} \\ \\
				- V_{n} &&  \pm h^{\fr12}(1 \pm V_{k}V^{k})
			\end{bmatrix}, \quad 
	        m^{MN}=h^{-\fr{1}{10}}
				\begin{bmatrix}
				h^{\fr12}(h^{mn}\pm V^mV^n) && \pm V^{m} \\ \\
				\pm V_{n} &&  \pm h^{-\fr12}
			\end{bmatrix}			
\end{aligned}
\end{equation}
with $V^m = \frac{1}{3!} \ve^{mnkl}C_{nkl}$ and $h=\det h_{mn}$. The variables $h_{mn}$ and $C_{mnk}$ will be later related to the components of the ordinary metric and the 3-form on the $d=4$ subspace.
Under generalised Lie derivatives the metric transforms as
\begin{equation}
\begin{aligned}
\mathcal{L}_{\L^{KL}}m_{MN} &=\fr12\L^{KL}\partial_{KL}m_{MN} + (\partial_{MK}\L^{LK})m_{LN} + (\partial_{NK}\L^{LK})m_{ML} \\&-\fr25(\partial_{KL}\L^{KL})m_{MN},
\end{aligned}
\end{equation}
that is as a generalised tensor of weight $\l[m_{MN}]=0$.

\def\kkappa{\gamma}

In order to obtain an explicit relationship between the 11-dimensional fields and those of the full SL(5) exceptional field theory one must perform the Kaluza-Klein decomposition under the $7+4$ split, and then rearrange the resulting fields into combinations covariant under the generalised Lie derivative of the SL(5) theory. Proceeding along these lines, we closely follow the $E_6$ discussion of \cite{Hohm:2013vpa} with minor changes relevant for the SL(5) group. One starts with the fields $h_{\m\n}$ and $h_{mn}$ which are related to the full $d=11$ metric by the usual Kaluza-Klein ansatz for the vielbein $E$ ($\hat\m, \hat a=1,\ldots 11$):
\begin{equation}
\label{KKgauge0}
  E_{\hat{\mu}}{}^{\hat{a}}  =  
\begin{bmatrix} h^{-\fr1{10}}h_{\mu}{}^{\bar{\mu}} & &
  A_{\mu}{}^{m} h_{m}{}^{\bar{m}} \\ \\ 0 & & h_{m}{}^{\bar{m}}
  \end{bmatrix}.
\end{equation}
Here $h_\m{}^{\bar{\m}} $ is the 7-dimensional vielbein defined as $h_{\m\n}=h_\m{}^{\bar{\m}}h_{\n}{}^{\bar{\n}}\h_{\bar{\m}\bar{\n}}$, $h_m{}^{\bar{m}}$ is the 4-dimensional vielbein defined as $h_{mn}=h_m{}^{\bar{m}} h_n{}^{\bar{n}} \h_{\bar{m}\bar{n}}$ and $h=\det{h_{mn}}$. Note the factor $h^{-\fr1{10}}$ that is needed to end up with the correct Einstein-Hilbert action in 7 dimensions. 

An important point is that the external metric is not a scalar under the generalised Lie derivative, since it transforms under its four-dimensional part $\L^m$ as
\begin{equation}
   \delta_{\Lambda}h_{\m\n} =  \Lambda^m\partial_m h_{\mu \n}+\fr25\, \partial_m\Lambda^m\,h_{\mu\n}.
\end{equation}
Thus, $h_{\m\n}$ is a weighted scalar of weight $\l[h_{\m\n}]=\fr25$. 
This is crucial for defining the procedure of deformation in analogy with the $d=10$ case as in \cite{Bakhmatov:2018bvp,Fernandez-Melgarejo:2018wpg}, namely as a rotation of the generalised metric by a specific matrix taking values in the duality group (SL(5) in our case). In order to perform this in a consistent manner, we rescale the generalised metric so as to absorb the degrees of freedom contained in $\det h_{\m\n}$. As explained below, this makes a connection between the full exceptional field theory defined above and its truncated version described in \cite{Blair:2013gqa}.
  
Similarly, one can apply the general prescription of the Kaluza-Klein reduction to the $d=11$ 3-form to obtain the following tower of $p$-forms:
  \begin{equation}
   \begin{split}
    A_{mnk} &= C_{mnk}, \\
    A_{\mu\,mn} &= C_{\mu mn}-A_{\mu}{}^k\,C_{kmn}, \\
    A_{\mu\nu\,m} &= C_{\mu\nu m}-2 A_{[\mu}{}^n\,C_{\nu]mn}+A_{\mu}{}^n A_{\nu}{}^k\,C_{mnk}, \\
    A_{\mu\nu\rho} &= C_{\mu\nu\rho}-3 A_{[\mu}{}^m\,C_{\nu\rho]m}+3 A_{[\mu}{}^m A_{\nu}{}^n\, C_{\rho]mn}
    -A_{\mu}{}^m A_{\nu}{}^n A_{\rho}{}^k\,C_{mnk}. 
   \end{split}
   \label{comp3form}
  \end{equation}
Note that in contrast to \cite{Hohm:2013vpa} we are using the conventions where the 3-form kinetic Lagrangian has the prefactor $-1/48$,
  \begin{equation}\label{3formterm}
  {\cal L}_{\text{\,3-form}} 
 =  -\frac{1}{ 48}E\,F^{\hat{\m}\hat{\n}\hat{\r}\hat{\s}}F_{\hat{\m}\hat{\n}\hat{\r}\hat{\s}}.
\end{equation}
This, together with the action for the Kaluza-Klein vector $A_\m{}^m$ following from the eleven-dimensional Einstein-Hilbert action produces kinetic terms for all $p$-forms in the theory. 

The ExFT generalised metric $m_{MN}$ contains the scalar degrees of freedom, which are encoded in the metric $h_{mn}$ and the gauge field $C_{mnk}$. The corresponding terms in the $d=11$ action read
\begin{equation}
    \begin{aligned}
    \mL_{sc}&=\mL_{\rm EH}-\fr{1}{48}e h^{-\fr15}h^{m p}h^{n q}h^{k r}h^{l s}F_{mnkl}F_{pqrs},
    \end{aligned}
\end{equation}
where $\mL_{EH}$ denotes the Lagrangian
 \begin{equation}
   \label{VEH}
    h^{\fr15} e^{-1}\mL_{\rm EH} (h,e) 
    = \mc{R}[h_{(4)}]+\frac{1}{4}h^{mn}\left(D_mh^{\mu\nu}\,D_{n}h_{\mu\nu}+h^{\m\n}D_mh_{\m\n}\,h^{\r\s}D_nh_{\r\s}\right),
   \end{equation} 
and we define the following combinations covariant under the internal diffeomorphisms
\begin{equation}
    D_{m} h_{\m\n} = \dt_{m}h_{\m\n} - \fr{1}{5}(h^{kl}\dt_{m}h_{kl}) h_{\m\n}.
\end{equation}
Substituting the explicit form of the generalised metric \eqref{mmetric1} it is straightforward to check that the above can be written in an SL(5) covariant form as \eqref{Lsc}. Note how the above expression differs from that of \cite{Blair:2014zba} in the part that includes only $m_{MN}$ and its derivatives. The reason is that the above reproduces the action with a prefactor of $h^{-\fr15}$, which follows from the proper Kaluza-Klein ansatz, rather than the action $\sqrt{h}(\mc{R}[h_{(4)}]-1/48 F^2)$ as one would expect in a truncated theory. More details on that in the following section.

Although by using the dualisation procedure it is possible to provide full identification between the 11-dimensional action and the SL(5) covariant action, for the purposes of the present paper we are not interested in topological terms of the SL(5) theory. Moreover, in the next section we will truncate the theory to describe only a special class of backgrounds that are relevant for our discussion.

\section{Equations of motion}
\label{section:eoms}

\subsection{Truncation to the extended space}

The general procedure for deforming a supergravity background in the ExFT/DFT formalism is based on switching between the geometric and non-geometric frames representation of the same generalised metric, and interpreting the non-geometric tri/bi-vector as a deformation parameter rather than a fundamental field~\cite{Bakhmatov:2018bvp,Bakhmatov:2019dow}. In this approach the deformation tensor can only include Killing vectors of the `internal' part of the background, using the terms of exceptional field theory. 

In order to simplify the discussion further, we consider only the backgrounds with the metric in a block-diagonal form, i.e.\ $M_{11}=M_4\times M_7$, where the internal metric $h_{mn}$ does not depend on the external coordinates $y^\m$. This allows to significantly simplify the equations of motion by truncating the theory to the purely scalar SL(5) extended geometry, similar to that of \cite{Berman:2010is, Blair:2014zba}, however keeping track of the external space geometry.  Taking this into account, the equations of motion following from the full SL(5) exceptional field theory are truncated to the case when
\begin{equation}
\label{ansatz}
    \begin{aligned}
       &h_{\m\n}=h_{\m\n}(y^\m, x^m), && m_{MN}=m_{MN}(x^m), \\
       &A_{\m}{}^{MN}=0, && B_{\m\n\, m} =0.
    \end{aligned}
\end{equation}
Moreover, given the structure of the theory, the second line above can be applied already at the Lagrangian level. This simplifies the exceptional field theory setup, leaving us with only the $d=7$ Einstein-Hilbert term and the scalar potential for the generalised metric in the action
\begin{equation}
\label{Ltrunc0}
\begin{aligned}
 e^{-1}\mL = & \ \mc{R}[h_{(7)}] - \bigg(\frac{1}{8}\,  {\partial}_{MN}{{m}_{P Q}}\,  {\partial}_{KL}{{m}^{P Q}}\,  {m}^{M K} {m}^{N L}      +\fr12 m^{MK}\dt_{MN}m^{NL}(h^{\m\n}\dt_{KL}h_{\m\n}) \\
 &+ \frac{1}{2}\, {\partial}_{MN}{{m}^{L N}}\,  {\partial}_{KL}{{m}^{M K}} +\frac{1}{2}\, {\partial}_{MN}{{m}_{P Q}}\,  {\partial}_{KL}{{m}^{M P}}\,  {m}^{N K} {m}^{L Q}\\ &+ \fr18 m^{MK}m^{NL}(h^{\m\n}\dt_{MN}h_{\m\n})(h^{\r\s}\dt_{KL}h_{\r\s})+\fr18 m^{MK}m^{NL}\dt_{MN}h^{\m\n}\dt_{KL}h_{\m\n}  \bigg),
\end{aligned}
\end{equation}
where $\mc{R}[h_{(7)}]$ is the Ricci curvature scalar of the metric $h_{\m\n}$. It is important to note, that such a truncation is background dependent, based on the specific ansatz \eqref{ansatz}, and does not provide a full consistent truncation of the theory. However, taking a specific initial solution of the form~\eqref{ansatz}, one is guaranteed to end up with a valid $d=11$ solution when making a tri-vector deformation, as long as the chosen Killing vectors do not introduce any dependence on the external coordinates $y^\m$. Note that although we truncate the Lagrangian, the structure of the couplings is such that the truncation at the level of equations of motion would be equivalent.

In what follows we will be interested in the case where a deformation results in rescaling of the 7-dimensional part of the metric by a single $x^m$-dependent factor. The $d=7$ metric before the deformation will be restricted to the form $h_{\m\n}(y^\m,x^m)=e^{-2\f(x^m)}h^{\fr15}\bar{h}_{\m\n}(y^\m)$, which allows to hide the $x^m$ dependence $\f(x^m)$ inside a properly rescaled generalised metric. To achieve this, define the rescaling as follows
\begin{equation}\label{rescaling}
    \begin{aligned}
     h_{\m\n} & = e^{-2 \f}h^{\fr15}\bar{h}_{\m\n},\\
     m_{MN} & = e^{-\f}h^{\fr1{10}} M_{MN}.
    \end{aligned}
\end{equation}
This implies that the Lagrangian $\mL = e \hat{\mc{R}}[h_{(7)}]+ \mL_{sc}$ can be rewritten as
\begin{equation}
\label{L_cov_truncated}
    \begin{aligned}
       \mL=&\ \bar{e}\,M^{-1} \left(\mc{R}[\bar{h}_{(7)}]- \frac{1}{8}\, {M}^{K L} {M}^{M N} {\dt}_{K M}{{M}_{P Q}}\,  {\dt}_{L N}{{M}^{P Q}} - \frac{1}{2}\,  {\dt}_{N K}{{M}^{M N} }\,  {\dt}_{M L}{{M}^{K L}}\right. \\
       &+ \frac{1}{2}\, {M}^{K L} {M}^{M N} {\dt}_{M K}{{M}^{P Q}}\,  {\dt}_{P L}{{M}_{N Q}}  + {M}^{K L} {M}^{M N} {\dt}_{K P}{{M}_{M N}}\,  {\dt}_{L Q}{{M}^{P Q}} \\&-\left.\frac{15}{24}\, {M}^{K L} {M}^{M N} {M}^{P Q} {M}^{R S} {\dt}_{M P}{{M}_{K L}}\,  {\dt}_{N Q}{{M}_{R S}}  \right),
    \end{aligned}
\end{equation}
where $M=\det M_{MN}= e^{5\f}h^{-1/2}$ and $\bar e = (\det \bar h_{\m\n})^{1/2}$. For the rescaling~\eqref{rescaling} the $d=11$ vielbein can be written in the following nice form
\begin{equation}
E_{\hat{\mu}}{}^{\hat{a}} \ = \ 
  \left(\begin{array}{cc} e^{-\f}\bar{e}_{\mu}{}^{a} &
  A_{\mu}{}^{m} h_{m}{}^{\alpha} \\ 0 & h_{m}{}^{\alpha}
  \end{array}\right),
\end{equation}
while the generalised metric becomes
\begin{equation}\label{mmetric}
\begin{aligned}
			M_{MN}=e^{\f}
				\begin{bmatrix}
				|h|^{-\fr12}h_{mn} && -V_{n} \\ \\
				- V_{m} &&  \pm |h|^{\fr12}(1\pm V_{k}V^{k})
			\end{bmatrix},
			\;\;
			M^{MN}=e^{-\f}
				\begin{bmatrix}
				|h|^{\fr12}(h^{mn}\pm V^mV^n) && \pm V_{n} \\ \\
				\pm  V_{m} &&  \pm |h|^{-\fr12}
			\end{bmatrix}	
\end{aligned}
\end{equation}
with $V^m= \frac{1}{3!}\, \ve^{mnkl}C_{nkl}$ and $h=\det h_{mn}$. Substituting this into \eqref{L_cov_truncated} one gets for the Lagrangian
\begin{equation}\label{lagr}
    \bar{e}^{-1}h^{-\fr12} \mL =  e^{-5\f} \mR[\bar{h}_{(7)}] +e^{-7\f}\left(\mR[h_{(4)}]+42 h^{mn}\dt_m \f\dt_n \f\mp\fr12 \nabla_{m}V^m\nabla_nV^n\right).
\end{equation}
Note that when $\mR[\bar{h}_{(7)}]=0$, the covariant Lagrangian \eqref{L_cov_truncated} reproduces the SL(5)$\times \RR^+$ Lagrangian of \cite{Blair:2014zba} up to full derivative terms.

\subsection{Deformation map}

The rescaled metric $M_{MN}$~\eqref{mmetric} can be written in terms of the generalised vielbein, $M_{MN} = \mE_M{}^A \h_{AB} \mE_N{}^B$, using
\begin{equation}
\label{O}
    \mE_M{}^{A} = e^{\fr{\f}{2}}
        \begin{bmatrix}
			|g|^{-1/4} g_m{}^{a} && |g|^{1/4} v^a \\ \\
			0 &&  |g|^{1/4}
		\end{bmatrix}, \quad	
	\h_{AB} =
		\begin{bmatrix}
			\h_{ab} && 0 \\ \\
			0 &&  1
		\end{bmatrix},
\end{equation}
where $g_m{}^a$ is a vielbein for the metric $g_{mn}$ and $g=\det g_{mn}$. 

Representation in terms of the vielbein proves to be the most convenient to define the deformations through an extension of the DFT $\b$-shift, which we call the $\W$-shift:
\begin{equation}
\label{def_map}
    \mE_M{}^A \longrightarrow O[\W]_M{}^N \mE_N{}^B, 
\end{equation}
with the matrix $O[\W]$ given by
\begin{equation}
    O[\W] = \begin{bmatrix}
      \d_m{}^n && 0 \\
      \\
      \fr1{3!}\e_{mpqr}\W^{pqr} && 1
    \end{bmatrix},
\end{equation}
where $\e_{mnkl}$ is the epsilon symbol and $\W^{mnk}$ are tensor components of the deformation tri-vector $\W = \fr{1}{3!} \r^{\a\b\g}\, k_\a\wedge k_\b\wedge k_\g$. The deformation in this form is completely frame independent and allows to define deformations for backgrounds with fluxes. 

Consider the initial background comprised by the internal metric $g_{mn}$, gauge 3-form field encoded by  $v^{m}= \frac{1}{3!}\, \ve^{mnkl}c_{nkl}$ and  the $7\times 7$ block of the 11-dimensional metric $g_{\m\n} = e^{-2\f(x)} \bar{g}_{\m\n}(y)$. The deformation map~\eqref{def_map} changes the corresponding generalised vielbein by a factor of $O[\W]$. The generalised metric acquires some $\W$ dependence; however, we are free to interpret the new metric as a standard expression~\eqref{mmetric}, only with some new deformed background fields $G_{mn}, V^m$ and $G_{\m\n}=e^{-2\Phi(x)} \bg_{\m\n}(y)$ (recall that we restrict ourselves to such deformations, where the external metric changes only by an $x^m$-dependent factor). Explicitly this can be written as
\begin{equation}\label{gen-metric-L}
\begin{aligned}
M_{MN} &= e^{\f}
			\begin{bmatrix}
                |g|^{-1/2}(g_{mn}\pm(1\pm v^2)W_mW_n-2v_{(m}W_{n)}) && -v_n\pm (1\pm v^2) W_n\\ \\
				-v_m \pm (1\pm v^2) W_m && \pm |g|^{1/2}(1 \pm v^2)
			\end{bmatrix}\\
 &= e^{\Phi}
\begin{bmatrix}
|G|^{-1/2}G_{mn} && -V_m \\ \\
-V_n && \pm|G|^{1/2}(1 \pm V^2)
\end{bmatrix},
\end{aligned}		
\end{equation}
where $W_m =\frac{1}{3!}\,\ve_{mnkl}\,\W^{nkl}$. The first matrix above is just the result of multiplication \eqref{def_map}, while the second matrix already contains the deformed fields. Equality between these two generalised metrics is what defines the deformation in terms of $d=11$ fields $(g_{mn},g_{\m\n},c_{mnk}) \longrightarrow (G_{mn},G_{\m\n},C_{mnk})$. Since the main transformation~\eqref{def_map} is essentially a frame change, it is convenient to refer to these two representations of the generalised metric as the $C$-frame and the $(C-\W)$-frame, depending on which fields appear inside the generalised metric.

In order to recover the explicit relations for  $d=11$ fields, we follow the same procedure as in \cite{Bakhmatov:2019dow} and start by equating determinants of the generalised metric in both frames~\eqref{gen-metric-L} to obtain
\begin{equation}\label{ext}
e^{5\Phi}|G|^{-\fr12}=e^{5\f}|g|^{-\fr12},
\end{equation}
where $G=\det G_{mn}$. Next, equating the generalised metrics  block-by-block one writes
\begin{subequations}\label{eqabc}
	\begin{align}
		e^{\Phi }|G|^{-\fr12}G_{mn} &= e^{\f }|g|^{-\fr12}  \left( g_{mn} \pm(1\pm v^2) W_m W_n - 2 v_{(m}W_{n)} \right), \label{eqa}\\
		e^{\Phi } V_m &= e^{\f } \Big(v_m{\pm}(1\pm v^2)W_m\Big). \label{eqc}
	\end{align}
\end{subequations}
Taking determinant of the first line and using the algebraic identity
\begin{equation}
\det \left( \d_m{}^n \pm (1\pm v^2) W_m W^n - v_m W^n - W_m v^n \right) = 1 \pm   W_m W^m - 2 W_m v^m + \big(W_m v^m \big)^2,
\end{equation}
we can define
\begin{equation}
    K^{-1} = e^{-6(\Phi - \f)}= 1 \pm   W_m W^m - 2 W_m v^m + \big(W_m v^m \big)^2.
\end{equation}
This gives the transformation rule $e^\Phi=K^{\fr16}e^\f$ for the field $\f$ and hence for the external metric. Understanding $K$ as a function of the deformation parameter $W_m$, the  equations in~\eqref{eqabc} express the deformed fields in terms of the original metric $g_{mn}$, gauge field $v^{m}$ and the deformation tensor $W_m$. Altogether, the deformation rules can be summarised as follows:
\begin{equation}\label{Cdef}
\begin{aligned}
        G_{\m\n} &= K^{-\fr13}g_{\m\n},\\
        G_{m n} & = K^{\fr23}\left( g_{mn} \pm(1\pm v^2) W_m W_n - 2 v_{(m}W_{n)} \right), \\
     	C^{mnk} & = K^{-1}\Big(c^{mnk}+(1\pm \fr1{3!}c^2) \W^{mnk}\Big).
\end{aligned}
\end{equation}
Note that the indices of $C_{mnk}$ are raised by the deformed metric $G_{mn}$, while the indices of $c_{mnk}$ are raised by the corresponding initial metric $g_{mn}$. It is worth reminding ourselves that the external $G_{\m\n}$ and the internal $G_{mn}$ blocks of the full $d=11$ metric are defined by the following interval
\begin{equation}
    ds^2=G_{\m\n}(y,x)dy^\m dy^\n+ G_{mn}(x)dx^m dx^n,
\end{equation}
and the external metric has the form $G_{\m\n}(y,x)=e^{-2\Phi(x)}\bar{g}_{\m\n}(y)$ for the initial $\bar{g}_{\m\n}$ that does not depend on the internal coordinates.

Observe that while the deformation prescription~\eqref{Cdef} is an extension of the recipe of~\cite{Bakhmatov:2019dow,Blair:2014zba}, it also reproduces the earlier results of~\cite{Bergshoeff:2000jn,Berman:2001rka}. In the latter context~\eqref{Cdef} is to be viewed as a field redefinition between the open and closed membrane frames in M-theory.

\subsection{Equations of motion}

Consider now the dynamical equations that control the deformation tensor $W_{m}$, given that the initial and the deformed backgrounds satisfy the equations of motion of the full $d=11$ supergravity, or equivalently of the truncated theory. 

For technical reasons we will consider the equations that govern deformations of the AdS${}_4\times \SS^7$ background in the $C$-frame, i.e., using the second form of the metric in~\eqref{gen-metric-L}. Hence, equations for the deformation tensor $W_{m}$ are implicit in this case. One starts with the Lagrangian of the truncated SL(5) ExFT in the $C$-frame~\eqref{lagr}
\begin{equation}
    \bar{e}^{-1}h^{-\fr12} \mL =  e^{-5\f} \mc{R}[\bar{h}_{(7)}] +e^{-7\f}\left(\mR[h_{(4)}]+42 h^{mn}\dt_m \f\dt_n \f+\fr12 \nabla_{m}V^m\nabla_nV^n\right).
\end{equation}
Equations of motion for the dynamical fields $\f, h_{mn}$ and $V_{m}$ then become
\begin{equation}
\label{eoms_V}
    \begin{aligned}
    &\d\f: &&  \fr57e^{2\f}\, \mc{R}[\bar{h}_{(7)}]+  \mR[h_{(4)}] + 12\, {\nabla}_{m}{{\nabla}_{n}{\f}\, }\,  {h}^{m n} - 42\, {\nabla}_{m}{\f}\,  {\nabla}_{n}{\f}\,  {h}^{m n} + \frac{1}{2}\, (\nabla V)^2=0, \\
    & \d V^{m}: && {\partial}_{m}(\nabla V) - 7\, (\nabla V) {\dt}_{m}{\f}=0, \\
    & \d h^{mn}: &&  {\mR}_{m n}[h_{(4)}]  - 7\, {\partial}_{m}{\f}\,  {\partial}_{n}{\f}\,  + 7\, {\nabla}_{m}{\nabla}_{n}{\f}\\
    & && +h_{m n}\left(-\fr12 e^{2\f}\mc{R}[\bar{h}_{(7)}]- \frac{1}{2}\, \mR[h_{(4)}] +28\, {\partial}_{k}{\f}\,  {\partial}_{l}{\f}\,  {h}^{k l}- 7\, {\nabla}_{k}{\nabla}_{l}{\f}\,  {h}^{k l} + \frac{1}{4}\, (\nabla V){}^{2} \right)=0,
    \end{aligned}
\end{equation}
These prove to be much simpler for further calculations than the original equations of motion of eleven-dimensional supergravity. The external space is always fixed to be the 7-sphere with the metric $\bg_{\m\n}$ up to a prefactor $e^{-2\f}$. Any supergravity solution of the form~\eqref{ansatz}, before or after the deformation, must also be a solution to these equations.

To derive explicit equations for the deformation tensor in the AdS${}_4\times \SS^7$ background, one would have to work in the mixed $(C-\W)$-frame using the generalised metric \eqref{gen-metric-L} in the Lagrangian \eqref{L_cov_truncated}. This provides formulation of eleven-dimensional supergravity in terms of both $C_{mnk}$ and $\W^{mnk}$, however, with the restriction that $\W^{mnk}$ is non-dynamical and rather encodes deformations. Given the complicated form of the generalised metric \eqref{gen-metric-L}, this appears to be a technically involved procedure and we will leave it beyond the scope of the present paper. Explicit construction of such a formulation for both DFT and ExFT is an open problem.

\section{\texorpdfstring{$\mathrm{AdS}_4 \times \SS^7$}{AdS4xS7} background}
\label{section:ads}

As an application of the developed formalism, let us look at the deformations of AdS$_4\times\SS^7$ spacetime. We will study the deformations that correspond to $\W\sim P\wedge P\wedge M$ and $D\wedge P\wedge P$. The fields of the initial eleven-dimensional solution  may be expressed as 
\begin{equation}
\textrm{d} s^2 = \frac{1}{4} \textrm{d} s^2 (\mathrm{AdS}_4) + R^2 d\W_{(7)}^2, \quad
F_4 = \frac{3}{8R} \text{vol}_{\mathrm{AdS}_4},  
\end{equation}
with a unit metric on the seven-sphere. We consider the AdS part as the `internal' space for the SL(5) ExFT. Denoting the AdS coordinates as $x^m=(x^0,x^1,x^2,z)$, the metric is as usual
\begin{equation}
    ds^2(\mathrm{AdS}_4)=\fr{R^2}{z^2}\left[-(dx^0)^2+(dx^1)^2+(dx^2)^2+(dz)^2\right].
\end{equation}
The only component of the flux and the corresponding 3-form gauge potential then become
\begin{equation}
\label{AdS_gauge_field}
\begin{aligned}
    F_{012z}&=-\fr{3R^3}{8z^4} , &&
    c_{012}&=-\fr{R^3}{8 z^3}.
\end{aligned}
\end{equation}

In this work we are interested in the tri-vector deformations of generalised Yang-Baxter type 
\begin{equation}
    \W= \frac{1}{3!}\, \r^{\a\b\g}k_\a \wedge k_\b \wedge k_\g,
\end{equation}
where $k_\a$ are Killing vectors of the initial background, in our case AdS$_4\times \SS^7$. As it has been mentioned in the previous section, the deformation matrix $O[\W]$ does not depend on the frame chosen, which implies that one may use Killing vectors of AdS${}_4$ in the $C$-frame\footnote{To arrive at the same conclusions one may follow arguments based on generalised Killing vectors of the initial undeformed backgrounds in the spirit of \cite{Sakamoto:2018krs}. }. Hence, we list Killing vectors of the AdS${}_4$ space:
\begin{equation}
\begin{aligned}
&&P_{a} &= \dt_a, & K_a &= x^2 \dt_{a} + 2 x_a D,\\
&&D&=-x^m\dt_m, &  M_{ab} &= x_a \dt_b - x_b \dt_a,
\end{aligned}
\end{equation}
where $a,b=0,1,2$ and $m,n=0,1,2,z$, and we define $x^2 = \h_{mn} x^m x^n$ and $x_a = \h_{ab}x^b$.

To proceed with explicit examples of deformed AdS${}_4\times \SS^7$ backgrounds systematically, we consider such combinations of the Killing vectors, that the resulting $\W$ is polynomial of order $0, 1$, etc.\ in powers of AdS coordinates. Applying the transformation rule \eqref{Cdef} we derive the deformed metrics $G_{\m\n}, G_{mn}$ and the 3-form $C_{mnk}$ from their undeformed initial values $g_{mn}, g_{\m\n}, c_{mnk}$ and the deformation tensor $W_{m}$ defined by the given choice of $\W^{mnk}$. To check whether a deformation gives a solution of equations of motion of 11-dimensional supergravity we substitute the deformed background written in terms of the fields $\Phi, G_{mn}, V^m$ into the equations of motion \eqref{eoms_V} of the truncated ExFT. Since the $\SS^7$ part only receives a correction encoded in the prefactor $e^{-2\f}$, using the truncated equation proves technically much simpler than the full $d=11$ theory.

\subsection{\texorpdfstring{$P\wedge P\wedge P$}{PPP  deformation}}

Start with the tri-vector as a polynomial of order 0 in coordinates, which corresponds to the trivial abelian $P\wedge P\wedge P$ deformation defined as
\begin{equation}
 \W = \frac{1}{3!}\,\r^{\a\b\g}k_\a\wedge k_\b \wedge k_\g= 4\h\, P_0 \wedge P_1 \wedge P_2.
\end{equation}
The deformation tensor and the prefactor $K$ then become
\begin{equation}
\begin{aligned}
W = -\frac{\h}{4} \fr{R^4}{z^4}dz, \quad K = \left( 1 + \h \fr{R^3}{z^3}\right)^{-1}.
\end{aligned}
\end{equation}
Following the prescribed procedure one finds for the deformed background
\begin{equation}
\label{PPP}
\begin{aligned}
ds^2 &= \fr{R^2}{4 z^2}\, \left( 1 + \h\fr{R^3}{z^3} \right)^{-\fr23} \left[ -(dx^0)^2 + (dx^1)^2 + (dx^2)^2 \right] +  
R^2\, \left( 1 + \h\fr{R^3}{z^3} \right)^{\fr13} \left( \fr{1}{4z^2}\,dz^2 + d\W_{(7)}^2 \right),\\
F &= -\fr38 \fr{R^3}{z^4}\,\left( 1 + \h \fr{R^3}{z^3} \right)^{-2}\,dx^0 \wedge dx^1 \wedge dx^2 \wedge dz,
\end{aligned}
\end{equation}
which is a solution of the equations \eqref{eoms_V} and hence of the $d=11$ equations of motion. 

For this deformation the Q-flux $Q_{m}{}^{nkl}=\dt_m \W^{nkl}$ can be checked to have no trace $Q_m{}^{mnk}=0$, hence the solution can be consistently reduced to a solution of the 10-dimensional type IIA theory. 
In fact, this $P\wedge P\wedge P$ deformation is abelian in the sense that there exists a generator that commutes with the other two. In the present case any of the $P_a$'s satisfies this property, say $P_2$. This implies that the deformation~\eqref{PPP} can be viewed as a result of dimensional reduction of the initial AdS${}_4\times \SS^7$ to IIA along $x^2$, then a TsT deformation with respect to the bi-vector $\b$, such that $\W=\dt_2\wedge \b$: 
\begin{equation}
\b = 4\h\, \dt_0 \wedge \dt_1,
\end{equation}
and finally an uplift back to $d=11$. As expected, this reflects the fact that the corresponding deformation is simply a $d=11$ extension of a TsT~\cite{CatalOzer:2009xd,Deger:2011nb}.

\subsection{\texorpdfstring{$P\wedge P\wedge M$}{ PPM deformation} }

The very next example with $\W$ being a polynomial of order 1 in $x^{a}$ provides a nonabelian deformation. Using the coefficients with the symmetry $\r^{ab,cd} = \r^{[ab],[cd]}$, consider
\begin{equation}
\label{PPM}
    \W = \frac14\, \r^{ab,cd} P_{a} \wedge P_{b} \wedge M_{cd} = \frac{4}{R^3}\,\r_a x^a\, \dt_0 \wedge \dt_1 \wedge \dt_2,
\end{equation}
where 
\begin{equation}
\label{rho_ppm}
\begin{aligned}
    \r_0 &= \frac{R^3}{4}(\r^{02,01} - \r^{01,02}),\\
    \r_1 &= \frac{R^3}{4}(\r^{01,12} - \r^{12,01}),\\
    \r_2 &= \frac{R^3}{4}(\r^{02,12} - \r^{12,02}),
\end{aligned}
\end{equation}
and we have introduced a numerical coefficient for convenience. It is easy to see that there is no such generator that commutes with all the others, which means that this deformation is non-abelian. The deformation tensor is
\begin{equation}
W = -\fr{R}{4z^4}\, \r_a x^a\, dz, \quad K = \frac{z^3}{z^3 - \r_a x^a},
\end{equation}
and the resulting deformed background then is given by
\begin{equation}\label{ppm}
\begin{aligned}
ds^2 &= \fr{R^2}{4} \left( z^3 - \r_a x^a \right)^{-\frac23} \left[ -(dx^0)^2 + (dx^1)^2 + (dx^2)^2 \right] +\fr{R^2}{z} \left( z^3 - \r_a x^a \right)^{\frac13} \left( \frac{1}{4z^2} dz^2 + d\W_{(7)}^2 \right),\\
F &= -\frac{3R}{8} \left(\fr{Rz}{ z^3 - \r_a x^a}\right)^2 dx^0 \wedge dx^1 \wedge dx^2 \wedge dz.
\end{aligned}
\end{equation}
Using \eqref{eoms_V} one can check that this provides a solution to 11-dimensional equations of motion for arbitrary values of the constants $\r_{a}$. In contrast to the previous example, trace of the Q-flux is non-zero and reads
\begin{equation}
2\dt_{[m} W_{n]}dx^m \wedge dx^n=-\fr{R^3}{4z^4}\r_a dx^a \wedge dz \neq 0.
\end{equation}
Upon dimensional reduction from ExFT in the $\W$-frame to $\beta$-supergravity one expects that non-vanishing trace $Q_m{}^{mkl}$ generates non-vanishing trace of the Q-flux of $\b$-supergravity. The latter is known \cite{Araujo:2017enj} to correspond to the vector $I$ of generalised supergravity.

\subsection{\texorpdfstring{$D\wedge P \wedge P$}{DPP deformation}}

Another way to build a tri-vector of the first order in powers of $x^{m}$ is to use the dilatation generator $D$ together with momenta. For the conformal algebra of AdS${}_4$ there are three possible pairs of $P_{a}, P_{b}$. It is convenient to parametrise a generic tri-vector of the form $D\wedge P \wedge P$ as
\begin{equation}
\label{dpp-omega}
    \W = \frac{2}{R^3}\, \r_{a} \e^{abc}\, D\wedge P_{b} \wedge P_{c} = \frac{4}{R^3}\, \r_{a} x^{a}\, \dt_0 \wedge \dt_1 \wedge \dt_2 - \frac{2}{R^3}\, z\, \r_{a} \e^{abc}\, \dt_{b} \wedge \dt_{c} \wedge \dt_z,
\end{equation}
with $\r_a$ corresponding to the three independent components of the $\r$-matrix. Using this $\W$, the deformation tensor and the prefactor are
\begin{equation}
    W = \frac{R}{4z^3}\, \r_a  \left( dx^a - x^a\, \frac{dz}{z} \right), \quad K = \left( 1+\dfrac{\r_a x^a}{z^3} - \dfrac{\r^2}{4z^4}\right)^{-1},
\end{equation}
where we define $\r^2=\r_{a}\r_{b}\h^{ab}$. The deformed background is then given by
\begin{equation}\label{dpp:sol}
\begin{aligned}
ds^2 &= \fr{R^2}{4} \left( z^3 + \r_a x^a - \frac{\r^2}{4z} \right)^{-\fr23} \left[-(dx^0)^2+(dx^1)^2+(dx^2)^2 +\left( 1+ \frac{\r_a x^a}{z^3} \right)\,dz^2 -\fr1{z^2}\r_a dx^a dz \right]\\
&+ \fr{R^2}{z}\left( z^3 + \r_a x^a - \frac{\r^2}{4z} \right)^{\fr13} d\W_{(7)}^2,\\
F &= -\fr{3 R^3 z^2}{8}  \left( 1+ \frac{\r^2}{12 z^4} \right) \left( z^3 + \r_a x^a - \frac{\r^2}{4z} \right)^{-2}\, dx^0 \wedge dx^1 \wedge dx^2 \wedge dz.
\end{aligned}
\end{equation}
By checking either~\eqref{eoms_V} or the field equations of $d=11$ supergravity one can show that this background is a solution, if parameters $\r_{a}$ form a null vector:
\begin{equation}\label{dpp}
    \r^2 = -\r_0^2+\r_1^2+\r_2^2 = 0.
\end{equation}
This is reminiscent of the $d=10$ Yang-Baxter deformation with $\Theta = \t^a M_{ab} \wedge P^b$, also parametrised by a null vector $\t$. The exact manner in which the condition~\eqref{dpp} arises is completely analogous to the way in which the Yang-Baxter equation is encoded in $d=10$ supergravity. One simply finds a factor of $\r^2$ out front every field equation after some simplifying algebra. We take this as a hint, that the condition~\eqref{dpp} may be an elementary example of a generalised Yang-Baxter equation, as applied to the tri-vector~\eqref{dpp-omega}.

Similar to the $P\wedge P\wedge M$ case, this background is an example of a deformation with vanishing R-flux, but non-vanishing trace of the $Q$-flux. For the latter one calculates 
\begin{equation}
2\dt_{[m} W_{n]}dx^m \wedge dx^n=-\fr{R}{z^4}\r_{a} dx^{a} \wedge dz \neq 0.
\end{equation}
Following the same arguments as in the previous subsection we conclude that the obtained deformed background cannot be reduced to a solution of conventional $d=10$ supergravity. Moreover, since the tri-vector $\W$ is non-ablelian the $P\wedge P\wedge M$ and $D\wedge P\wedge P$ deformations cannot be put to the form $\W=\dt_* \wedge \b$. The conclusion is that both these solutions are proper 11-dimensional deformations that cannot be accessed via 10-dimensional techniques.

An important question is what fraction of supersymmetry is preserved by the tri-vector deformation. We note that only half of the $AdS$ Killing spinors are invariant under the spinorial Lie derivative~\cite{Kosmann:Dec1971,Figueroa-OFarrill:1999klq} with respect to shifts, $\mathcal{L}_{P_i} \e = 0$. We expect therefore that the $P\wedge P\wedge P$ and $P\wedge P\wedge M$ solutions of the present article have their supersymmetries halved by the deformation. Moreover, the dilatations break even these remaining Killing spinors of~$AdS$, $\mathcal{L}_D \e = \frac12 \e$, which presumably makes the $D\wedge P\wedge P$ deformed solution non-supersymmetric. Preservation of supersymmetry in the bi-vector deformation case  was the subject of a recent investigation~\cite{Orlando:2018kms,Orlando:2018qaq,Orlando:2019his}, which has produced a closed form expression for the Killing spinors after the deformation in terms of the bi-vector parameter $\Th$. Implementing this for the tri-vector deformations in full generality deserves a detailed separate study.

\subsection{\texorpdfstring{$D\wedge K\wedge K$}{DKK deformation}}

The outer automorphism of the conformal algebra
\begin{equation}
    P_a \longleftrightarrow K_a,\quad D\longleftrightarrow -D
\end{equation}
can be realised geometrically by an inversion, which is an isometry of $AdS$ spacetime:
\begin{equation}
x^a \longrightarrow \frac{x^a}{x^2 + z^2},\quad z \longrightarrow \frac{z}{x^2 + z^2}. 
\end{equation}
Applying this map to the $D\wedge P\wedge P$-deformed background~\eqref{dpp:sol}, one should be able to recover the deformation with $\Omega \sim D \wedge K \wedge K$. Given the geometric symmetry, one expects this $D\wedge K\wedge K$ deformation to also be a solution. Note that the tri-vectors are in fact very closely related,
\begin{equation}
    D \wedge K_a \wedge K_b = (x^2 + z^2)^2 \, D \wedge P_a \wedge P_b.
\end{equation}
Explicit calculation shows, however, that already the second equation in \eqref{eoms_V}, which states $\nabla_m V^m e^{-7\f}$ $=\text{const}$, does not hold for the obtained background. This negative result makes it very intriguing to derive explicit equations for the deformation tensor, that is the equations of motion \eqref{eoms_V} in the mixed $(C-\W)$-frame, and investigate the reason of such unexpected behaviour more closely.

\section{Conclusions and discussions}
\label{section:conclusions}

In this work we studied tri-vector deformations of the AdS${}_4\times \SS^7$ solution of 11-dimensional supergravity, generalising the results of \cite{Bakhmatov:2019dow} to the case of non-abelian deformations. Working in the formalism of SL(5) exceptional field theory properly truncated to describe backgrounds of the form $M_4\times M_7$, we generalise the deformation map of \cite{Bakhmatov:2019dow} to the case of backgrounds with non-vanishing 3-form flux and provide two examples of non-abelian deformations. The corresponding tri-vector deformation parameter is schematically given by  $\W\sim P\wedge P \wedge M$ and $\W \sim D \wedge P \wedge P$, where $D, P_a, M_{ab}$ stand for generators of the AdS${}_4$ symmetry algebra. Both deformations are non-abelian, that is one cannot represent the tri-vector in the form $\W=\dt_* \wedge \b$ where $\dt_*$ commutes with the generators of $\b$. The deformed backgrounds cannot be obtained by reducing to 10 dimensions, performing a bi-vector deformation and uplifting back to $d=11$, as there is no obvious direction for the reduction (see e.g.~\cite{Deger:2011nb}). 

Our proposed procedure may be used to further investigate the AdS${}_4\times \SS^7$ solution in search for more non-abelian deformations, as well as to address the deformations of the sphere part of AdS${}_7\times \SS^4$. The isometry algebra of a sphere may turn out to be too restrictive, though. Indeed, the only deformations of $d=10$ supergravity backgrounds found so far are either abelian or of the so-called Jordanian type. The latter means that the generators chosen for the Killing bi-vector deformation ansatz belong to the Borel subalgebra of the full isometry algebra. So are the tri-vector deformations $D\wedge P \wedge P $ and $M \wedge P \wedge P$ discussed in this paper, which may be referred to as the tri-vector deformations of generalised Jordanian type. It is an interesting question, whether these correspond to solutions of a generalisation of the Yang-Baxter equation. Based on the known examples in $d=10$ and $d=11$ and on the analysis of the equations of motion, one can expect that the algebraic equations that restrict a deformation to be a solution must be quadratic in $\r^{abc}$. By an appropriate choice of basis for the algebra such constraints can be brought to the form $\r^I \r^J \k_{IJ}=0$, where $\k_{IJ}$ is some invariant tensor and $I,J$ are some (multi)indices corresponding to the chosen basis. In the examples of this paper the Borel subalgebra contains in particular the generators $P_0,P_1,P_2$ with the symmetry group $SO(1,2)$ and the invariant tensor becomes just the Minkowski metric $\h_{ab}$. This may be viewed as a motivation of the non-trivial equation \eqref{dpp}. For $SO(d+1)$, which is the symmetry group of $\SS^{d}$ with the Euclidean metric $\d_{IJ}$ as its invariant tensor, we  expect an equation of the type $\sum (\r^I)^2=0$, which has only a trivial solution. This has already been observed in~\cite{Bakhmatov:2019dow}, where no nontrivial non-abelian bi-vector deformations of the flat Euclidean space were found.

Part of the motivation for constructing the non-abelian tri-vector deformations was to test the proposals for generalised Yang-Baxter equation that have appeared recently. In \cite{Bakhmatov:2018apn} it has been shown using techniques of Double Field Theory and $\b$-supergravity that for a bi-vector deformation $\b=\fr12 r^{\a\b}k_\a \wedge k_\b$ to generate a solution to the field equations of $d=10$ supergravity, it is sufficient that the matrix $r^{\a\b}$  satisfy the classical Yang-Baxter equation. The same condition is imposed by assuming that the R-flux vanishes. Turning to M-theory backgrounds one naturally considers tri-vector instead of bi-vector. In~\cite{Bakhmatov:2019dow} the vanishing of the ExFT R-flux $R^{m,nklp} = \W^{mq[n} \dt_q \W^{klp]}$ was proposed as the condition for a tri-vector deformation to be a solution. Assuming the tri-Killing ansatz for $\W$~\eqref{3k}, $R=0$ translates into 
\begin{equation}
\label{R-flux=0}
6 \r^{\a\b[\g} \r^{\d\e|\z|} f_{\a\z}{}^{\h]} + \r^{[\g\d\e} \r^{\h]\a\z} f_{\a\z}{}^{\b} = 0.
\end{equation}
Explicit check shows that for the $P\wedge P\wedge M$ and $D\wedge P\wedge P$ deformations the R-flux indeed vanishes. However, at least for the $D\wedge P\wedge P$ this is not sufficient to end up with a solution to $d=11$ equations of motion, and a stronger algebraic constraint on $\r^{\a\b\g}$~\eqref{dpp} is required. 

Based on the generalisation of Poisson-Lie T-duality to the U-duality setup~\cite{Sakatani:2019zrs,Sakatani:2020iad}, an algebraic constraint for $\r^{\a\b\g}$ was proposed that was conjectured to be a sufficient condition for the deformation to be a supergravity solution. The non-abelian deformed solutions described in the present work are in the non-unimodular class, meaning $\dt_m \W^{mnk}\neq 0$, therefore the corresponding $\r^{\a\b\g}$ cannot satisfy the equations of \cite{Sakatani:2019zrs} as the latter suppose unimodularity. It is then natural to expect that the algebraic constraints for the tri-vector components $\r^{\a\b\g}$, such as~\eqref{dpp}, are manifestations of the M-theory generalisation of the CYBE with non-unimodularity properly taken into account. The Exceptional Drinfeld Algebra construction of~\cite{Malek:2019xrf} includes non-unimodular terms and may turn out to be the way that leads to the correct generalisation of the CYBE.  Note that while in the $d=10$ case both unimodular and non-unimodular deformations are required to satisfy the same classical Yang-Baxter equation, this seems not to be the case for M-theory. Moreover, the condition of the vanishing R-flux, which is equivalent to the CYBE in $d=10$, appears to be only a part of the equations of \cite{Sakatani:2019zrs}. 

Given these results, searching for the general algebraic equations for $\r^{\a\b\g}$ that generalise the classical Yang-Baxter equation appears to be an interesting direction of further research. From the algebraic point of view a natural expectation is that the CYBE, which is relevant for the scattering of particles in $1+1$ dimensions, will be promoted to the tetrahedron equation describing scattering of strings in $d=1+2$~\cite{zamolodchikov1981,frenkel1991}. Depending on the labeling scheme, the tetrahedron equation may be referred to as Zamolodchikov or Frenkel-Moore equation. Deriving a representation independent form of the semi-classical limit of the tetrahedron equation and comparing the results to those of~\cite{Sakatani:2019zrs} is an open problem.

More transparent is the algebraic interpretation of the vanishing R-flux condition. Following~\cite{Bagger:2006sk, Bagger:2007jr} one notices that the M2-brane world-volume dynamics brings about a non-commutativity parameter given by a tri-vector $\W^{mnk}$, as well as the following Nambu-Poisson 3-bracket
\begin{equation}
\{x^m,x^n,x^k\}=\W^{mnk}.
\end{equation}
The fundamental identity for such a bracket,
\begin{equation}
\begin{split}
\{\{x^i,x^j,x^k\}, x^l,x^m\} &-\{\{x^i,x^l,x^m\},x^j,x^k\}\\ &-\{x^i,\{x^j,x^l,x^m\},x^k\}-\{x^i,x^j,\{x^k,x^l,x^m\}\}=0,
\end{split}
\end{equation}
which is at the same time the closure condition for the Exceptional Drinfeld Algebra~\cite{Malek:2019xrf}, is precisely the vanishing R-flux condition of the SL(5) theory. Indeed,  when written in terms of $W_m =\frac{1}{3!}\,\ve_{mnkl}\W^{nkl}$ the fundamental identity is proportional to $\ve^{mnkl}W_{[n}\dt_k W_{l]}=0$, that is $R^{m,ijkl}\ve_{ijkl}=0$. Given this observation and the fact that all particular examples of tri-vector deformations are R-fluxless, it is reasonable to conjecture that any sensible M-theory background must have vanishing R-flux.

In this article our attention was focused on the $d=11$ supergravity deformations and underlying symmetries. On a more practical note, however, one may look into the holographic interpretations of the deformations that were obtained here. The dual field theory description of the AdS${}_4\times\SS^7$ supergravity background is given by the ABJM theory~\cite{Aharony:2008ug}, while some deformations thereof should correspond to the solutions that we have presented. The abelian deformation $P\wedge P \wedge P$ considered in this paper is a tri-vector analogue of the Maldcena-Russo deformation of AdS${}_5\times \SS^5$~\cite{Maldacena:1999mh}. In the field theory language this is represented by a non-commutative gauge theory whose product can be recovered either following the brane picture (see e.g.\ \cite{Imeroni:2008cr}), or by considering Drinfeld twists of the Hopf algebra structure of the corresponding tensor algebra as in~\cite{vanTongeren:2015uha}. In general, for any abelian tri-vector deformations along the AdS${}_4$ isometries one should expect non-commutative structures defined by the standard Moyal star product to appear in the ABJM theory~\cite{Martin:2017nhg}. More interesting are the non-abelian deformations $D\wedge P \wedge P$ and $M \wedge P \wedge P$, which can be understood as tri-vector generalisations of the so-called jordanian Yang-Baxter deformations. To gain some understanding of these deformations on the gauge theory side one could take the approach of~\cite{vanTongeren:2015uha}, extending it to what for want of a better name can be termed the ``exceptional Drinfeld twist''. This should define a twist of the matrix $\r^{abc}$ such that the tetrahedron equation is satisfied. Authors are not aware whether and to what extent such structures have been considered in the mathematical literature.

As a final remark we notice that in contrast to the approach of \cite{Bakhmatov:2018bvp}, in the present work we did not derive explicit equations for the deformation tensor $\W^{mnk}$ from exceptional field theory, rather working in the $C$-frame. The dynamical differential equations for $\W$ seem to be the optimal starting point for deriving algebraic constraints for the deformation parameters $\r^{\a\b\g}$. However, to address backgrounds with fluxes one should go to the mixed $(C-\W)$-frame, which we leave for future work.

\section*{Acknowledgements}
 
We thank S.~Deger, E.~Malek, E.~\'O~Colg\'ain, D.~Osten, Y.~Sakatani, M.M.Sheikh-Jabbari, and A.~Tseytlin for discussions and helpful comments on the final manuscript.
The work of IB and EtM is supported by the Foundation for the Advancement of Theoretical Physics and Mathematics ``BASIS'' and by the Russian Government program of competitive growth of Kazan Federal University. The work of EtM and KG is supported by Russian Ministry of education and science (Project 5-100).
 
\appendix

\section{Notations and conventions}
\label{section:app}

In this paper we use the following conventions for indices
\begin{equation}
    \begin{aligned}
       &\hat{\mu}, \hat{\nu}, = 1 \dots 11&& \mbox{eleven directions, curved}; \\
       &\hat{\alpha}, \hat{\beta}, = 1 \dots 11&& \mbox{eleven directions, flat}; \\       
       &\mu, \nu, \rho, \ldots = 1 \dots 7&& \mbox{external seven  directions, curved}; \\
       &\bar{\mu}, \bar{\nu}, \bar{\rho}, \ldots = 1 \dots 7&& \mbox{external seven  directions, flat}; \\       
       & m, n, k, l, \ldots = 1,\dots,4 && \mbox{internal four directions, curved}; \\
       & \bar{m}, \bar{n}, \bar{k}, \bar{l}, \ldots = 1,\dots,4 && \mbox{internal four directions, flat}; \\       
       & M, N, K, L, \ldots = 1,\dots, 5 && \mbox{fundamental ExFT indices, curved};  \\
       & A, B, C, D, \ldots = 1,\dots, 5 && \mbox{fundamental ExFT indices, flat};\\
       & \a, \b, \g, \ldots = 1,\dots, N && \mbox{indices labelling Killing vectors}; \\
       & a, b, c, d, \ldots = 0,\dots,2 && \mbox{first three directions of AdS${}_4$ in Poincar\'e patch}.\\
    \end{aligned}
\end{equation}
Totally antisymmetric tensor in $n$ dimensions is defined as
\begin{equation}
    \ve_{i_1\ldots i_n} = g^{1/2} \e_{i_1\ldots i_n},\qquad \e_{1\ldots n} = 1.
\end{equation}
Curvature tensors are defined as
\begin{equation}
\begin{aligned}[]
[\nabla_m,\nabla_n]V^k&= R_{mn}{}^k{}_lV^l, \\
R_{m n}{}^{k}{}_l & = \partial_{m}{\Gamma_{n l}{}^{k}} - \partial_{n}{\Gamma_{m l}{}^{k}} + \Gamma_{m q}{}^{k}\Gamma_{n l}{}^{q}-\Gamma_{n q}^{k}\Gamma_{m l}^{q},\\
R_{mn}&=R_{k m}{}^{k}{}_{n}.
\end{aligned}
\end{equation}
In our notations  non-vanishing commutators of the AdS algebra read
\begin{equation}
\begin{aligned}{}
[D,P_{a}]&=P_{a}, && & [D, K_{a}]&=-K_{a},\\
[M_{ab},P_{c}]&=-2\h_{c[a}P_{b]}, && & [M_{ab},K_{c}]&=-2\h_{c[a}K_{b]},\\
[P_{a},K_{b}]&=2M_{ab}+2 \h_{ab}D, && & [M_{ab},M_{cd}]&=-2\h_{c[a}M_{b]d}+2\h_{d[a}M_{b]c}.
\end{aligned}
\end{equation}
These can be mapped to standard commutation relations of so(2,3) algebra by defining
\begin{equation}
\begin{aligned}
J_{ab}&=iM_{ab}, && & J_{0*}&=iD,\\
J_{*a}&=\fr i2(P_{a}-K_{a}), && & J_{0a}&=\fr i2 (P_{a}+K_{a}).
\end{aligned}
\end{equation}
 
%\bibliographystyle{JHEP}
%\bibliography{biblio.bib}

\providecommand{\href}[2]{#2}\begingroup\raggedright\endgroup

\end{document}